\documentclass[it,preprint]{iucrmy}        % DO NOT DELETE THIS LINE

\journalcode{I}

\usepackage{graphicx}
\usepackage{color}
\usepackage{url}

\begin{document}                  % DO NOT DELETE THIS LINE

     %-------------------------------------------------------------------------
     % The introductory (header) part of the chapter
     %-------------------------------------------------------------------------

     % The title of the chapter. Use \shorttitle to indicate an abbreviated
     % title for use in running heads (you will need to uncomment it).

\title{EDA: EXAFS Data Analysis software package}
\shorttitle{EDA software package}

     % Authors' names and addresses.
     % Use \author for all authors' names.

     \author{Alexei}{Kuzmin}

     %\author{A. N.}{Author}

     % Use \shortauthor to indicate an abbreviated author list for use in
     % running heads (you will need to uncomment it).

\shortauthor{Kuzmin}

\maketitle                        % DO NOT DELETE THIS LINE

     %-------------------------------------------------------------------------
     % The main body of the chapter
     %-------------------------------------------------------------------------
     % Now enter the text of the document in multiple \section's, \subsection's,
     % \subsubsection's, \sectioniv's and \sectionv's as required.
\begin{abstract}

The EXAFS data analysis software package EDA consists of a suite of programs running under Windows operating system environment and designed to perform all steps of conventional EXAFS data analysis such as the extraction of the XANES/EXAFS parts of the x-ray absorption coefficient, the Fourier filtering, the EXAFS fitting using the Gaussian and cumulant models. Besides, the package includes two advanced approaches, which allow one to reconstruct the radial distribution function (RDF) from EXAFS based on the regularization-like method and to calculate configurational-averaged EXAFS using a set of atomic configurations obtained from molecular dynamics or Monte Carlo simulations. 
\end{abstract}

\section{General concept}

Below the EXAFS data analysis software package EDA \cite{EDA} is described in details emphasising its key features. The full package, documentation and application examples are available for download at \url{http://www.dragon.lv/eda/}.

The EDA package has been under continuous development from 1988. It has been created with an idea to be intuitively simple and fast, guiding the user step by step through each part of the EXAFS analysis.
Originally developed for the MS-DOS compatible operating systems, the current version of the package consists of a set (Table~\ref{table1}) of interactive programs running under Windows operating system environment. The originality of the EDA package is mainly related to (i) the procedure of the EXAFS oscillation extraction from the experimental data performed by the EDAEES code, (ii) the regularization-like method for the radial distribution function (RDF) reconstruction from EXAFS by the EDARDF code \cite{EDARDF} and (iii) the calculation of configuration-averaged EXAFS based on the results of molecular dynamics or Monte Carlo simulations by the EDACA code \cite{Kuzmin2009}.

\begin{figure}
\label{fig1}
\includegraphics[width=0.95\linewidth]{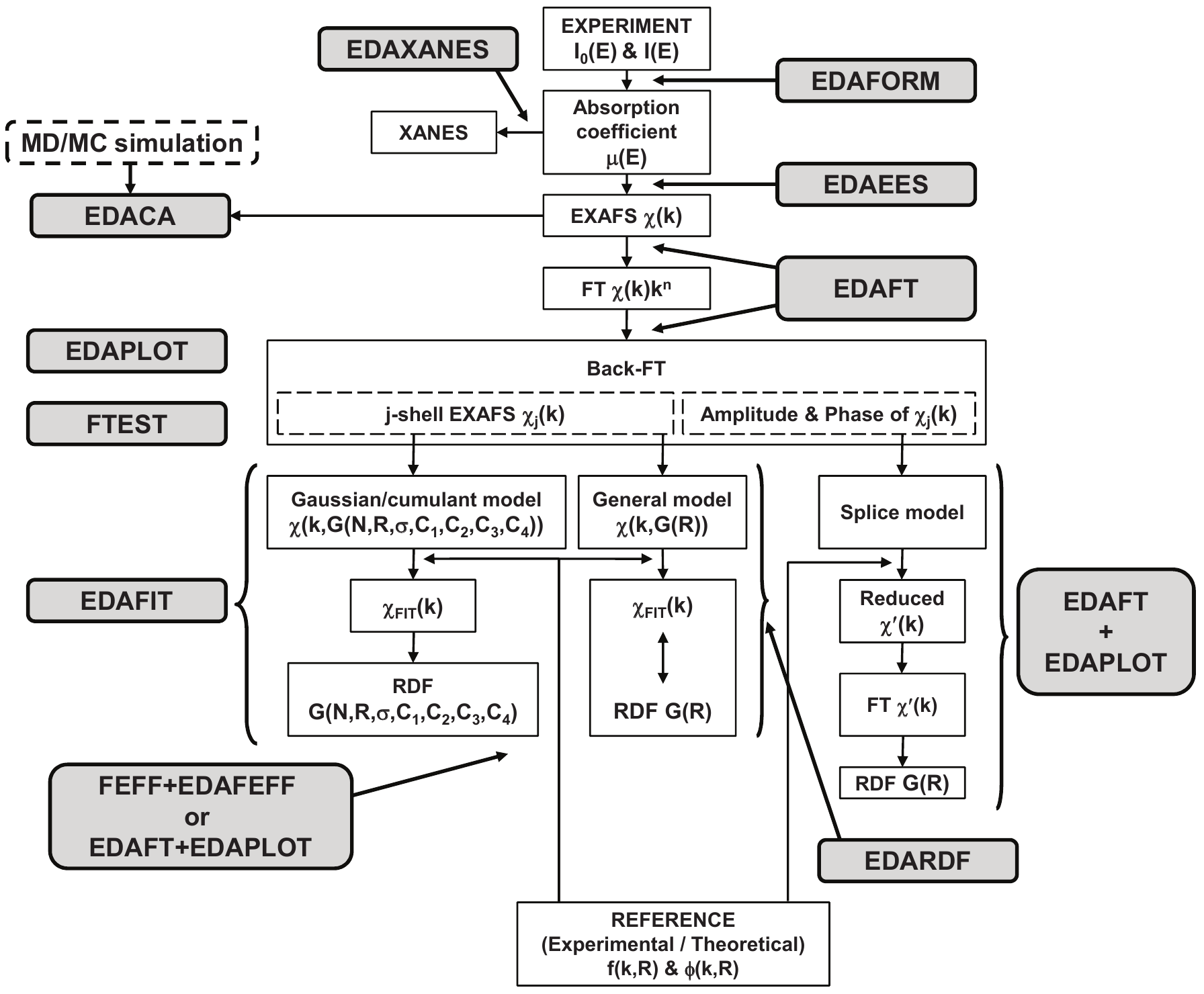}
\caption{Flowchart of the EXAFS data analysis by the EDA package.}
\end{figure}

The various components of the EDA package and their relation are shown in Fig.~\ref{fig1}, where the main steps of the analysis and the computer codes involved are given. We will describe it briefly below.

When performing XAS experiment, one usually obtains two signals $I_0(E)$ and $I(E)$, which are proportional to the intensity of the X-ray beam with the energy $E$ before and after interaction with the sample. These two signals are used to calculated the X-ray absorption coefficient $\mu(E)$ by the EDAFORM code.

At this point, the X-ray absorption near edge structure (XANES) of $\mu(E)$ can be isolated and
analysed by calculating its first and second derivatives using the EDAXANES code. This step allows one to determine  precisely the position of the absorption edge and, thus, to check the reproducibility of the energy scale for the single sample or to determine the absorption edge shift $\Delta E$ due to a variation of the effective charge of the absorbing atom in different compounds. 

The extraction of the EXAFS  $\chi(E) = (\mu(E) - \mu_b(E) - \mu_0(E))/ \Delta \mu_0(E)$ is implemented in the EDAEES code using the following sophisticated procedure. The background contribution $\mu_b(E)$ is determined  by extrapolating the pre-edge background as $\mu_b(E) = A - B/E^3$. Next, the atomic-like contribution $\mu_0(E)$ is determined  as
$\mu_0(E) = \mu_0^I(E) + \mu_0^{II}(k) + \mu_0^{III}(k)$, where the three functions $\mu_0^I(E)= P_n(E)$,
$\mu_0^{II}(k)= P_m(k)$ ($P_n$ is the polynomial of n-order) and $\mu_0^{III}(k)= S_3(k,p)$ ($S_3(k,p)$
is the smoothing cubic spline with the parameter $p$) are calculated in series, the first one in $E$-space
whereas the last two  in $k$-space ($k$ is the photoelectron wavenumber). The EXAFS normalization is performed as $\Delta \mu_0(E) = \mu_0^I(E)$.  Such procedure guarantees accurate determination of the $\mu_0(E)$ function, and as a result, the EXAFS $\chi(k)$, even if experimental data are far from ideal.
The $k$ scale is conventionally defined as $k=\sqrt{(2 m_e /\hbar^2)(E-E_0)}$, where $m_e$ is the electron mass, $\hbar$ is the Planck's constant,
and $E_0$ is the threshold energy, i.e., the energy of a free electron with zero momentum.
Deglitching and normalization of the EXAFS to the edge jump $\Delta \mu_0(E)$ obtained from the reference compound, theoretical tables \cite{Teo1986} or equal to a constant are also possible.

The extracted experimental EXAFS $\chi(k)$ can be directly compared with the
configuration-averaged EXAFS calculated based on the results of molecular dynamics (MD) or Monte Carlo (MC) simulations
by the EDACA code or can be analysed in more conventional way. In the latter case,
the Fourier filtering procedure (i.e. direct and back Fourier
transforms (FTs) with some suitable "window"-function)  is applied using the EDAFT code
to separate a contribution from the required range in $R$-space into the total EXAFS.
Such approach allows one to simplify the analysis, at least, for a contribution from the
first coordination shell of the absorbing atom.

Finally, the EXAFS from a single or several coordination shells can be simulated using
different models to extract structural information. The EDA package
allows one to use three models (the first two will be discussed below):
(i) conventional multi-component
parameterized model within the Gaussian or cumulant approximation
(the EDAFIT code, see Section~\ref{model1}), (ii) arbitrary RDF
model determined by the regularization-like approach (the EDARDF code, see Section~\ref{model2}),
(iii) the so-called "splice" model \cite{stern92} (combined use of the EDAFT and
EDAPLOT codes is required). To perform simulations, one needs to
provide the scattering amplitude $f(k,R)$\   and phase shift $\phi(k,R)$\ functions for each
scattering path. These data can be obtained from experimental
EXAFS spectrum of some reference compound, taken from tables \cite{Teo1986,McKale1988}
or calculated theoretically. In the EDA package, one has possibility to use the
theoretical data calculated by the FEFF8/9 codes \cite{FEFF8,FEFF9}, which can
be extracted from the feff****.dat files by the EDAFEFF code.

Finally, the EDAPLOT code is provided for visualization, comparison and
simple mathematical analysis of any data obtained within the EDA package.
Note that since all data are kept in simple ASCII format, they can be
easily imported to and treated by any other codes.

\section{Multi-component model within the Gaussian/cumulant approximation.}\label{model1}

The fitting of the EXAFS $\chi(k)$ in the $k$-space within the single-scattering curved-wave approximation is implemented in the EDAFIT code which is based on  the cumulant expansion of the EXAFS equation \cite{Rehr2000,Kuzmin2014IUCR}
\begin{eqnarray}
 \chi(k)
  &=&  \sum_{i}^{shells} S_0^2  {{N_i}\over{k R^2_i}} f_i(\pi,k,R_i)
 \exp(-2 \sigma_i^2 k^2 + {2 \over 3} C_{4i} k^4  \nonumber \\
 & - & {4 \over 45} C_{6i} k^6) \exp(-2 R_i / \lambda(k))
 \sin(2kR_i - {4 \over 3} C_{3i} k^3 \nonumber \\
 &+& {4 \over 15} C_{5i} k^5 + \phi_i(\pi,k,R_i))
\label{eq1}
\end{eqnarray}
where $k = \sqrt {k^{\prime 2} + (2m_e / \hbar^2) \Delta E_{0i}}$\ is the
photoelectron wavenumber corrected for the difference $\Delta
E_{0i}$\ in the energy origin between experiment and theory;
$S_{0}^2$\  is the scale factor taking into account amplitude
damping due to multielectron effects; $N_i$\  is the
coordination number of the i-th shell; $R_i$\  is the radius of the
i-th shell; $\sigma_i$\  is the mean-square relative displacement (MSRD) or
Debye-Waller factor; $C_{3i}$, $C_{4i}$, $C_{5i}$\ and $C_{6i}$\ are
cumulants of a distribution taking into account anharmonic effects
and/or non-Gaussian disorder; $\lambda(k) = k / \Gamma$\  ($\Gamma$\
is a constant) is the mean free path (MFP) of the photoelectron;
$f(\pi,k,R_i)$\  is the backscattering amplitude of the
photoelectron due to the atoms of the i-th shell;
$\phi(\pi,k,R_i) = \psi(\pi,k,R_i) + 2\delta_l(k) - l\pi$\ is the
phase shift containing contributions from the absorber
$2\delta_l(k)$\ and the backscatterer $\psi(\pi,k,R_i)$\ ($l$\ is
the angular momentum of the photoelectron).

The fitting parameters of
the model are $S_0^2N_i$, $R_i$, $\sigma^2_i$, $\Delta E_{0i}$, $C_{3i}$,
$C_{4i}$, $C_{5i}$, $C_{6i}$\ and $\Gamma$.  The maximum number of
fitting parameters, which can be used in the EXAFS model, is limited by
the Nyquist criterion $N_{\rm par} = 2 \Delta k \Delta R / \pi $ \cite{Stern1993}.

Note that when the functions $f(\pi,k,R_i)$\  and $\phi(\pi,k,R_i)$\
are  extracted from the EXAFS spectrum of a reference compound,
the values of fitting parameters will be {\it relative}.
To compare different models obtained by fitting of the
EXAFS using the EDAFIT code, Fisher's $F_{0.95}$\ criterion,
implemented in the FTEST code \cite{EDA}, can be applied.

\section{Regularization-like method}\label{model2}

The regularization-like method implemented in the EDARDF code allows one to determine {\it model
independent} RDF $G(R)$ from the experimental EXAFS.
It is especially suitable for the analysis of the first coordination shell in
locally distorted or disordered materials, such as low-symmetry crystals (e.g. with Jahn-Teller distortions), amorphous compounds, glasses, and systems with strongly anharmonic behaviour, where a decomposition into the cumulant series fails.
At the same time, the  method can be also used in more simple cases as a starting point for the selection
of a conventional model described in the previous section.

The RDF $G(R)$ is determined by inversion of the EXAFS equation within the single-scattering approximation
\begin{equation}
 \chi(k)
  =  S_0^2 \int_{R_{\rm {min}}}^{R_{\rm {max}}} {{G(R)}\over{k R^2}}
  f(\pi,k,R)  \sin(2kR  + \phi(\pi,k,R)) dR
\label{eq103}
\end{equation}
using the iterative method described in \cite{Kuzmin2000}. Two regularizing criteria
are applied after each iteration to restrict the shape of $G(R)$ to physically significant solutions: it must be positive-defined and smooth function. 

\begin{figure}
	\label{fig2}
	\includegraphics[width=0.80\linewidth]{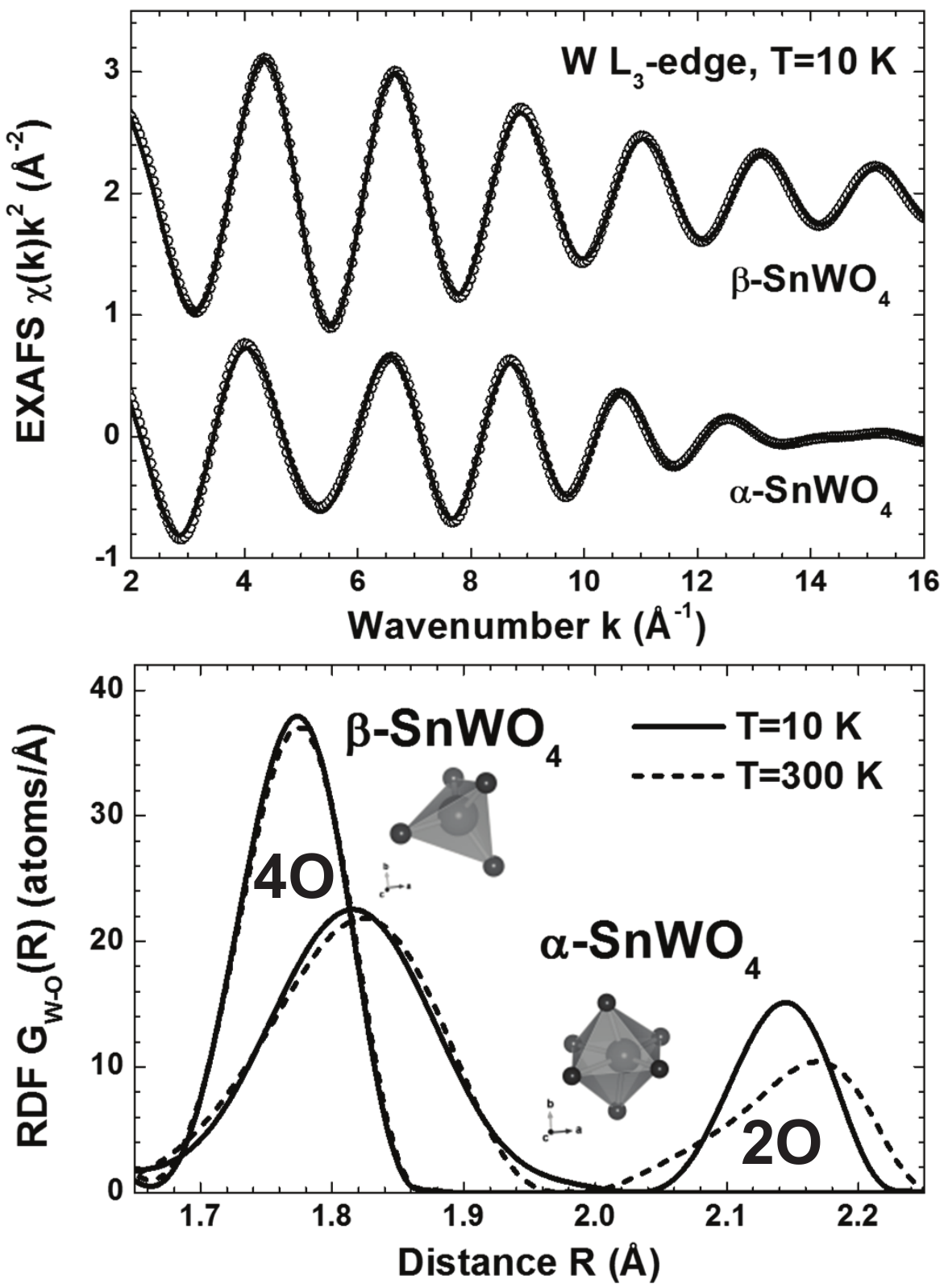}
	\caption{Upper panel: comparison of the experimental (circles) and calculated (solid lines) W L$_3$-edge EXAFS spectra $\chi(k)k^2$ for the first coordination shell of tungsten in 
		$\alpha$-SnWO$_4$ (lower curves) and $\beta$-SnWO$_4$ 
		(upper curves)  at 10~K.
		Lower panel: calculated radial distribution functions (RDF's) $G_{\rm W-O}(R)$
		for W--O bonds within the first coordination shell of tungsten in $\alpha$-SnWO$_4$ and $\beta$-SnWO$_4$ at 10~K (solid lines) and 300~K (dashed lines). The two groups of 4 and 2 oxygen atoms are indicated.}
\end{figure}

The use of the method is demonstrated in Fig.~\ref{fig2} for the case of tin tungstate, which exists in two phases -- $\alpha$-SnWO$_4$ and $\beta$-SnWO$_4$ \cite{Kuzmin2015}.
In orthorhombic $\alpha$-SnWO$_4$ phase tungsten atoms are six-fold coordinated by oxygen atoms, and the WO$_6$ octahedra are strongly distorted due to the second-order Jahn-Teller effect  because of the W$^{6+}$(5$d^0$) electronic configuration. The six W--O bonds in
$\alpha$-SnWO$_4$ are split into two groups of four short bonds at $\sim$1.82~\AA\ and two long bonds at $\sim$2.15~\AA. 
In cubic $\beta$-SnWO$_4$  tungsten atoms have slightly deformed WO$_4$ tetrahedral coordination with the W--O bond lengths of about 1.77~\AA. An increase of temperature from 10~K to 300~K affects weakly
the W--O bonding in the WO$_4$ tetrahedra and also the group of four shortest  W--O bonds in the WO$_6$ octahedra. At the same time, 
the distant group of weakly bound two oxygen atoms in the WO$_6$ octahedra shifts to longer distances and becomes more broadened. Thus, the reconstructed RDFs reproduce nicely the W L$_3$-edge EXAFS in both tin tungstates and allow one to follow a distortion of the tungsten first shell in details.

Another example, shown in Fig.~\ref{fig2}, is concerned the local atomic structure relaxation upon crystallite size reduction in ZnWO$_4$ \cite{Kalinko2011}. Crystalline ZnWO$_4$\ has monoclinic ($P2/c$) wolframite-type structure built up of distorted WO$_6$\ and ZnO$_6$\  octahedra joined by edges into infinite zigzag chains. 
The distortion of the metal-oxygen octahedra leads to
the splitting of the W--O and Zn--O distances  into three
groups of two oxygen atoms each with the bond lengths of about 1.79, 1.91 and 2.13~\AA\ around tungsten atoms and 2.03, 2.09 and 2.22~\AA\ around zinc atoms. Note that the three 
W--O groups are well resolved in the RDF. 
Upon the reduction of crystallite size down to $\sim$2~nm, 
significant relaxation of the atomic structure occurs leading to some broadening of 
the RDF peaks, especially at large distances (2.1-2.4~\AA), whereas the nearest oxygen atoms become stronger bound. Such structural changes in ZnWO$_4$ nanoparticles correlate with their optical and vibrational properties. 

Other application examples of the method include dehydration process in molybdenum oxide hydrate \cite{Kuzmin2000},
the effect of composition and crystallite size reduction in tungstates \cite{Kuzmin2001,Anspoks2014b,Kuzmin2014}, and studies of the local environment in glasses \cite{Kuzmin1997a,Rocca1998,Rocca1999,Kuzmin2006}.

\section{Configuration-averaged EXAFS simulations}

A particular feature of the EDA package is its ability to perform more advanced calculations of the configuration-averaged EXAFS
based on the results of MD simulations (Fig.~\ref{fig4}) \cite{Kuzmin2009,Kuzmin2014IUCR,Kuzmin2016val}. Note that a set of atomic configurations generated using the MC simulation \cite{PIMC2008} can be also used in a similar manner. 
To use this approach, called MD-EXAFS, one needs to provide an *.XYZ file containing temporal snapshots of atomic coordinates, which can be obtained from most MD codes such as GULP \cite{Gale2003}, DLPOLY \cite{DLPOLY}, LAMMPS \cite{LAMMPS1995} or CP2K \cite{VandeVondele2005}. Additionally the input file with a set of commands for the FEFF8/9 code is also required. 

A care should be taken to get proper configuration-averaged EXAFS. This means that 
a number of snapshots should be sufficiently large (usually few thousands) to get good statistics, and the time step between subsequent snapshots should be enough small to sample properly the dynamic properties of the material. The MD-EXAFS approach allows one to validate 
different theoretical models, e.g. force-fields, and/or perform the EXAFS interpretation 
far beyond the nearest coordination shells.

Application examples of this method cover several compounds: SrTiO$_3$ \cite{Kuzmin2009},  ReO$_3$ \cite{Kalinko2009jpcs}, Ge \cite{Timoshenko2011}, NiO \cite{Anspoks2010,Anspoks2012,Anspoks2014}, LaCoO$_3$ \cite{Kuzmin2011},
ZnO \cite{Timoshenko2014zno}, AWO$_4$ (A=Ca, Sr, Ba) tungstates \cite{Kalinko2016},  Y$_2$O$_3$ \cite{Jonane2016y2o3}, FeF$_3$ \cite{Jonane2016fef3} and UO$_2$ \cite{Bocharov2017}.

The case of microcrystalline and nanocrystaline (6~nm) NiO \cite{Anspoks2012} is illustrated in  Fig.~\ref{fig5}. The Ni K-edge EXAFS spectra of both compounds are dominated by a contribution from the first two coordination shells (Ni--O$_1$ and Ni--Ni$_2$) of nickel. However, the outer shells are responsible for a number of well resolved peaks located above $\sim$3~\AA\ in the Fourier transforms. Due to cubic rock-salt structure of NiO, the multiple-scattering events play an important role in the formation of EXAFS and must be treated properly. The classical MD simulations \cite{Anspoks2012} were performed using the  force-field model included two-body central force interactions between atoms
described by a sum of the Buckingham and Coulomb potentials. The effects of crystallite size, thermal disorder and Ni vacancy concentration were taken into account. The calculated configuration-averaged EXAFS reproduces well the experimental data for both nickel oxides. In the case of nanocrystalline NiO, the damping of the EXAFS oscillations due to the atomic structure relaxation and the progressive decrease of the FT peaks amplitude at longer distances are observed as a result of the crystallite size reduction. 

\ack{This work was supported by the Latvian Science Council grant no. 187/2012.}

\section{References}
%\referencelist
     %-------------------------------------------------------------------------
     % The back matter of the paper - references
     %-------------------------------------------------------------------------

%
%     %-------------------------------------------------------------------------
%     % TABLES AND FIGURES SHOULD BE INSERTED AFTER THE MAIN BODY OF THE TEXT
%     %-------------------------------------------------------------------------
%
%     % Simple tables should use the tabular environment according to this
%     % model
%
%\begin{table}
%\caption{Caption to table}
%\begin{tabular}{llcr}      % Alignment for each cell: l=left, c=center, r=right
% HEADING    & FOR        & EACH       & COLUMN     \\
%\hline
% entry      & entry      & entry      & entry      \\
% entry      & entry      & entry      & entry      \\
% entry      & entry      & entry      & entry      \\
%\end{tabular}
%\end{table}
%
%     % Postscript figures can be included with multiple figure blocks
%

     %-------------------------------------------------------------------------
     % The back matter of the paper - references
     %-------------------------------------------------------------------------

     %-------------------------------------------------------------------------
     % TABLES AND FIGURES SHOULD BE INSERTED AFTER THE MAIN BODY OF THE TEXT
     %-------------------------------------------------------------------------

     % Simple tables should use the tabular environment according to this
     % model

\begin{table}\label{table1}
\footnotesize
\caption{A set of programs for EXAFS data analysis and simulations included in the EDA software package. }
\begin{tabular}{ll}      % Alignment for each cell: l=left, c=center, r=right
\hline
 Code title& Code description    \\
\hline
EDAFORM &converts original experimental data from several beamlines into \\
        &the EDA file format (ASCII, 2 columns). \\
EDAXANES&extracts the XANES part of the experimental X-ray absorption\\
        &spectrum and calculates its first and second derivatives. \\
EDAEES &extracts the EXAFS part $\chi(k)$ using original algorithm for the\\
       &atomic-like ("zero-line") background removal. \\
EDAFT  &performs Fourier filtering procedure (direct and back Fourier\\
       &transforms) with or without amplitude/phase correction using\\
       &a number of different (rectangular, Gaussian, Kaiser-Bessel, \\
       &Hamming and Norton-Beer F3) window functions. \\
EDAFIT &is a non-linear least-squares fitting code, based on a high speed  \\
       &algorithm without matrix inversion. A multi-shell Gaussian or\\
       &cumulant model within the single-scattering approximation can \\
       &contain up to 20 shells with up to 8 fitting parameters ($N_i$,  $S_o^2$,   \\
       &$R_i$,  $\sigma_i^2$, $\Delta E_{0i}$, $C_{3i}$, $C_{4i}$, $C_{5i}$, $C_{6i}$) in each.
        Constrains on the range \\
       &of any fitting parameter or its value can be imposed. \\
EDARDF &is the regularization-like least-squares-fitting code allowing one\\
       &to determine model-independent RDF in the first coordination  \\
       &shell for a compound with arbitrary  degree of disorder. \\
FTEST  &performs analysis of variance of the fitting results based on\\
       &the Fisher's $F_{0.95}$-test. \\
EDAPLOT&is a general-purpose program for plotting, comparison, and \\
       &mathematical calculations frequently used in the EXAFS data \\
       &analysis (more than 20 different operations). \\
EDAFEFF&extracts the scattering amplitude and phase shift functions from \\
       &FEFF****.dat files, calculated by the FEFF8/9 code, for the use\\
       &with the EDAFIT or EDARDF codes. \\
EDACA  &calculates configuration-averaged EXAFS based on the results \\
       &of molecular dynamics simulations.\\
\hline
\end{tabular}
\end{table}

     % Postscript figures can be included with multiple figure blocks

\begin{figure}
\label{fig3}
\includegraphics[width=0.80\linewidth]{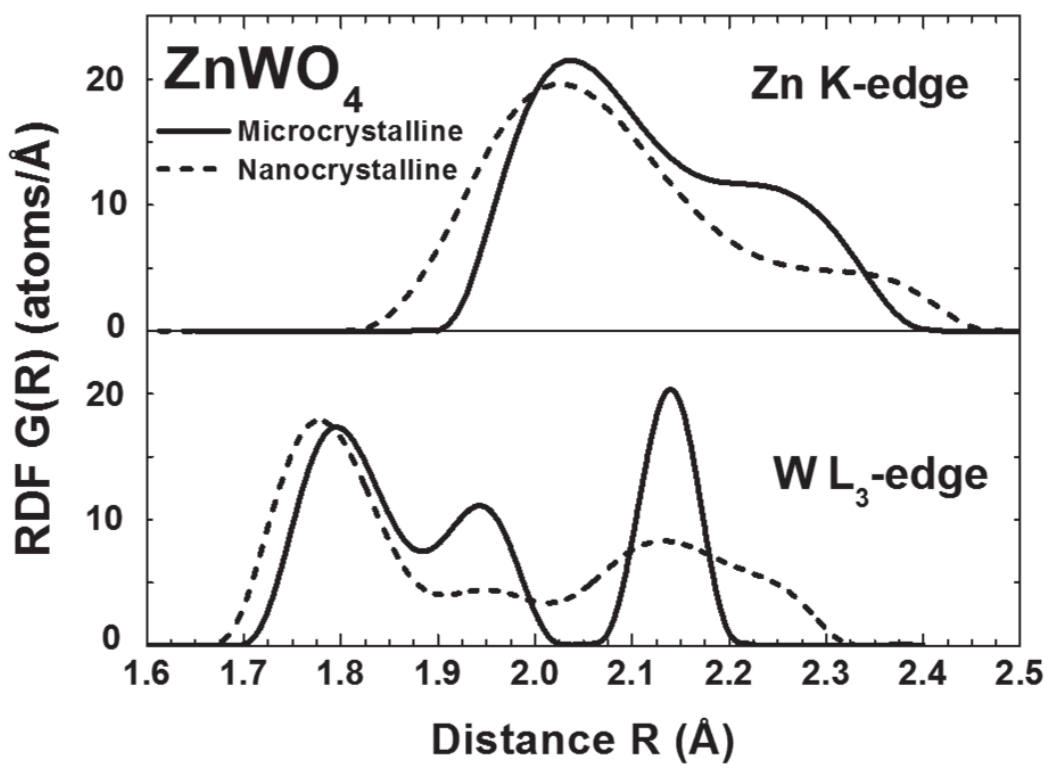}
\caption{The reconstructed RDFs $G(R)$ for the	first coordination shell of tungsten and zinc in nanoparticles and microcrystalline ZnWO$_4$.}
\end{figure}

\begin{figure}
	\label{fig4}
	\includegraphics[width=0.95\linewidth]{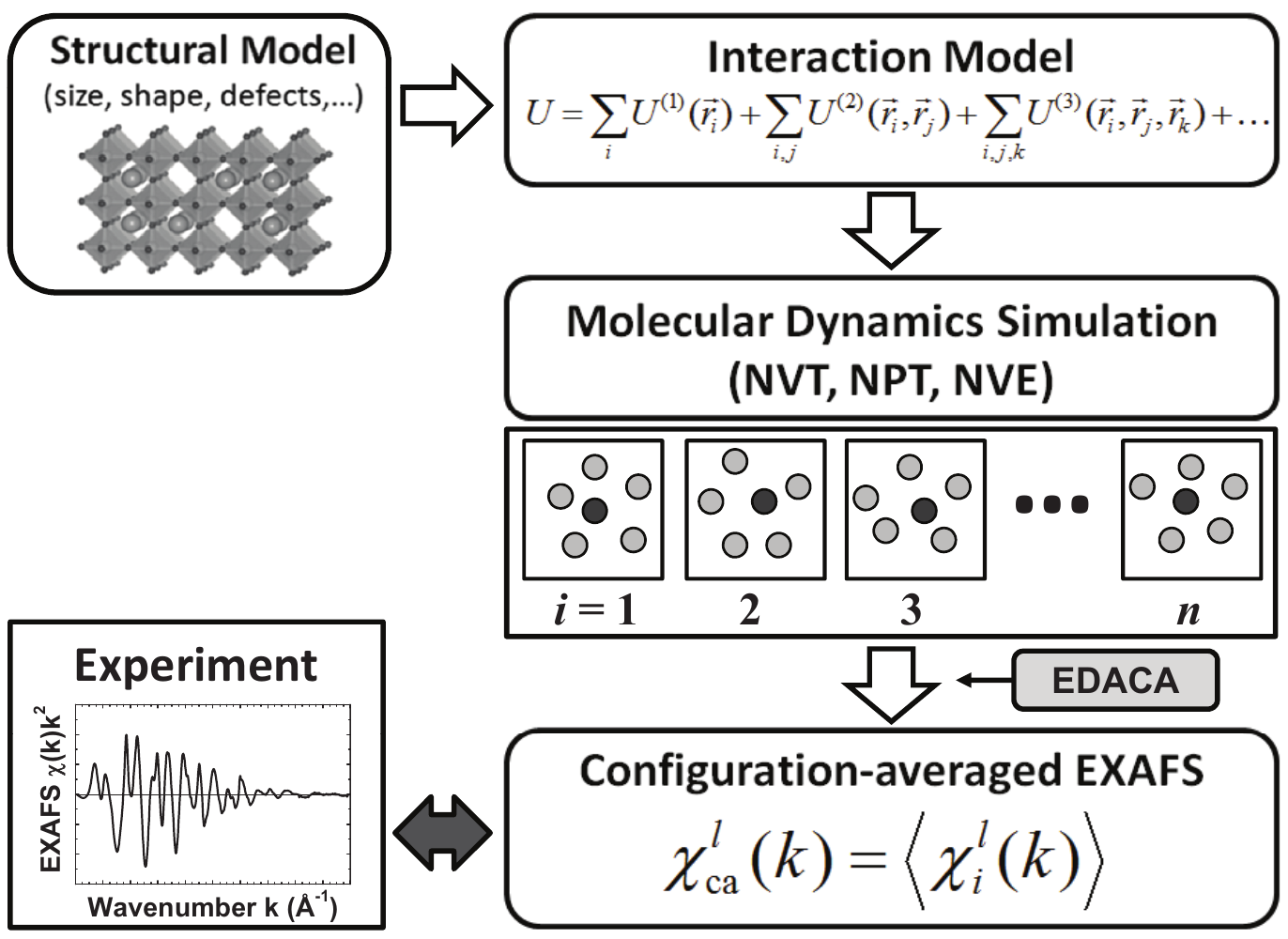}
	\caption{Flowchart of the MD-EXAFS calculations.}
\end{figure}

\begin{figure}
	\label{fig5}
	\includegraphics[width=0.80\linewidth]{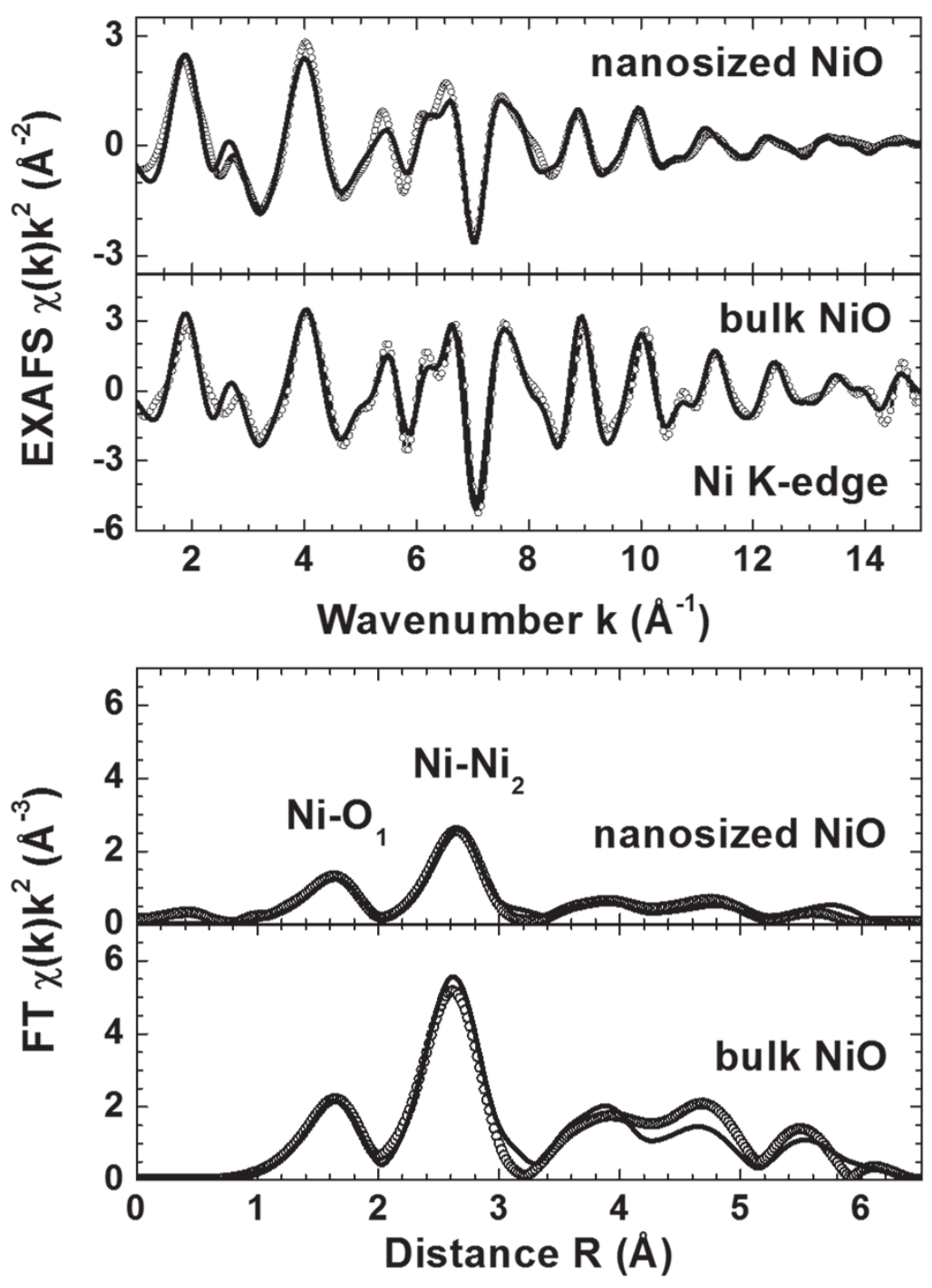}
	\caption{Comparison of the experimental (circles) and calculated configuration-averaged  (solid lines) Ni K-edge EXAFS spectra $\chi(k)k^2$ (upper panel) and their Fourier transforms (lower panel) for bulk and nanosized NiO at 300~K. }
\end{figure}

\twocolumn

\end{document}